\documentclass[ 
 aip,
 amsmath,amssymb,
reprint,%
]{revtex4-1}
\usepackage{graphicx}
\usepackage{dcolumn}
\usepackage{bm}

\usepackage[utf8]{inputenc}
\usepackage[T1]{fontenc}
\usepackage{mathptmx}
\usepackage{etoolbox}

\makeatletter
\def\@email#1#2{%
 \endgroup
 \patchcmd{\titleblock@produce}
  {\frontmatter@RRAPformat}
  {\frontmatter@RRAPformat{\produce@RRAP{*#1\href{mailto:#2}{#2}}}\frontmatter@RRAPformat}
  {}{}
}%
\makeatother

\renewcommand{\selectlanguage}[1]{}

\begin{document}

\preprint{AIP/123-QED}

\title[]{Applications of machine learning in ion beam analysis of materials}
\author{Tiago F. Silva}
\email{tfsilva@if.usp.br}
\affiliation{Instituto de Física da Universidade de S\~ao Paulo\\Rua do matão, trav. R 187, 05508-090 S\~ao Paulo, Brazil.}

\date{\today}

\begin{abstract}
Ion Beam Analysis (IBA) is an established tool for material characterization, providing precise information on elemental composition, depth profiles, and structural information in the region near the surface of materials. However, traditional data processing methods can be slow and computationally intensive, limiting the efficiency and speed of the analysis. This article explores the current landscape of applying Machine Learning Algorithms (MLA) in the field of IBA, demonstrating the immense potential to optimize and accelerate processes. We present how ML has been employed to extract valuable insights from large datasets, automate repetitive tasks, and enhance the interpretability of results, with practical examples of applications in various IBA techniques, such as RBS, PIXE, and others. Finally, perspectives on using MLA to approach open problems in IBA are also discussed.
\end{abstract}


\maketitle

\section{\label{intro}Introduction}

    Ion beam analysis (IBA) consists of a set of well-established analytical techniques that exploit interactions of swift ion beams (with a kinetic energy typically of the order of hundreds of keV up to tens of MeV) with matter to obtain elemental composition, depth profiles and structural information in the region near the surface of materials \cite{jeynes_ion_2011}. When combined with scanning systems, IBA produces two- or three-dimensional elemental maps with minimal or negligible damage to the sample. Its applications span a wide range of fields, including biology \cite{grime_high-throughput_2020}, life sciences \cite{whitlow_dynamic_2024}, sustainable energy \cite{mayer_ion_2019, rubel_role_2016, komander_accurate_2024}, forensics \cite{simon_addressing_2024}, planetary geology \cite{campbell_re-examination_2019, sargent_emulation_2021}, amd micro- and nanoelectronics \cite{laricchiuta_rutherford_2019, claessens_ensemble_2022, claessens_probing_2023}, among others.
    
    The key advantages of IBA include its potential as a primary standard for thin film char\-ac\-ter\-i\-za\-tion \cite{jeynes_rbs_2017}, traceability of uncertainties \cite{jeynes_quality_2006, jeynes_accurate_2012, colaux_high_2014, cureatz_improving_2022}, and the synergistic integration between the many signals into a consistent and accurate sample characterization \cite{jeynes_total_2012, jeynes_accuracy_2020, silva_self-consistent_2021, silva_bias_2022}.

    However, despite the maturity of its data processing workflows, which are typically based on simulation software with the most up-to-date physics models \cite{mayer_improved_2014, barradas_advanced_2008, rauhala_status_2006, campbell_guelph_2021}, IBA faces certain limitations. For example, reliance on a reverse Monte Carlo approach \cite{mcgreevy_reverse_1988} - where the description of the sample is obtained from the search for the best possible agreement between the experimental and simulated spectra - can result in slow convergence due to the high computational costs of the simulations, thus limiting throughput.

    Parallel to that, we are surrounded by systems employing Machine Learning Algorithms (MLA) for various purposes, from tailored advertisements based on our online experience to massive information processing for complex text composition. The reason behind this explosive growth is that MLA systems can identify patterns, make accurate predictions, and automate repetitive tasks using large amounts of data. This technology enables processes to be optimized by simplifying workflows and reducing human error, ultimately enhancing efficiency. MLA can extract valuable insights, make data-driven decisions, and improve overall productivity, making it a vital tool in today's rapidly evolving science.

    MLA has played a pivotal role in the advancement of scientific development in various disciplines. For example, MLA has been used to accelerate the discovery of new materials \cite{pyzer-knapp_accelerating_2022, szymanski_autonomous_2023}, and has generated valuable insights in nuclear science \cite{agency_artificial_2022}. However, their potential in IBA still needs improvement and can profoundly change the data processing workflow. One concept spread in the community is the multi-objective optimization applied in data fusion of different ion beam techniques \cite{jeynes_total_2012, jeynes_accuracy_2020, silva_multisimnra:_2016, silva_self-consistent_2021,  silva_bias_2022}. Furthermore, some other works have been done in the past showing the great potential of MLA, but the algorithm applied is restricted to models of artificial neural networks to fit experimental spectrometry data \cite{barradas_artificial_2000, vieira_neural_2000, barradas_artificial_2002}. Although these works demonstrated the potential to extend the applications of IBA methods \cite{pinho_artificial_2005, nene_artificial_2006, demeulemeester_artificial_2010}, the idea did not evolve much for more than a decade. Only recently, it came back with a boost in artificial intelligence and more diversity in the application of algorithms.

    For example, we apply different models in our laboratory to process our data qualitatively and quantitatively. We succeed in implementing unsupervised MLA for feature extraction in semi-automatic processing of wide-field PIXE mapping \cite{silva_multivariate_2018}, with the advantage of enabling disentanglement of pigment composition even in mixtures or layered structures. The algorithm extracts meaningful insights that are easily overlooked in conventional data processing. Our last development focuses on increasing the interpretability of neural network predictions when analyzing experimental data \cite{oliveira_what_2022}. With our method, we can identify the regions of the spectra that most significantly contribute to the output of the neural network. Methods like this are vital in a material analysis laboratory, mainly if some uncertainty traceability is requested.
    
    In this paper, I summarize the current status of MLA application in IBA and demonstrate what kind of benefits we may have by embracing this technology. The aim of this publication is to serve as a concise guide to those interested in applying those models in their data processing workflow, or to those interested in taking the state-of-the-art even further.
    
\section{\label{methods}Methods}

\subsection{More about IBA}

\begin{figure*}[htb!]
    \centering
    \includegraphics[width=\textwidth]{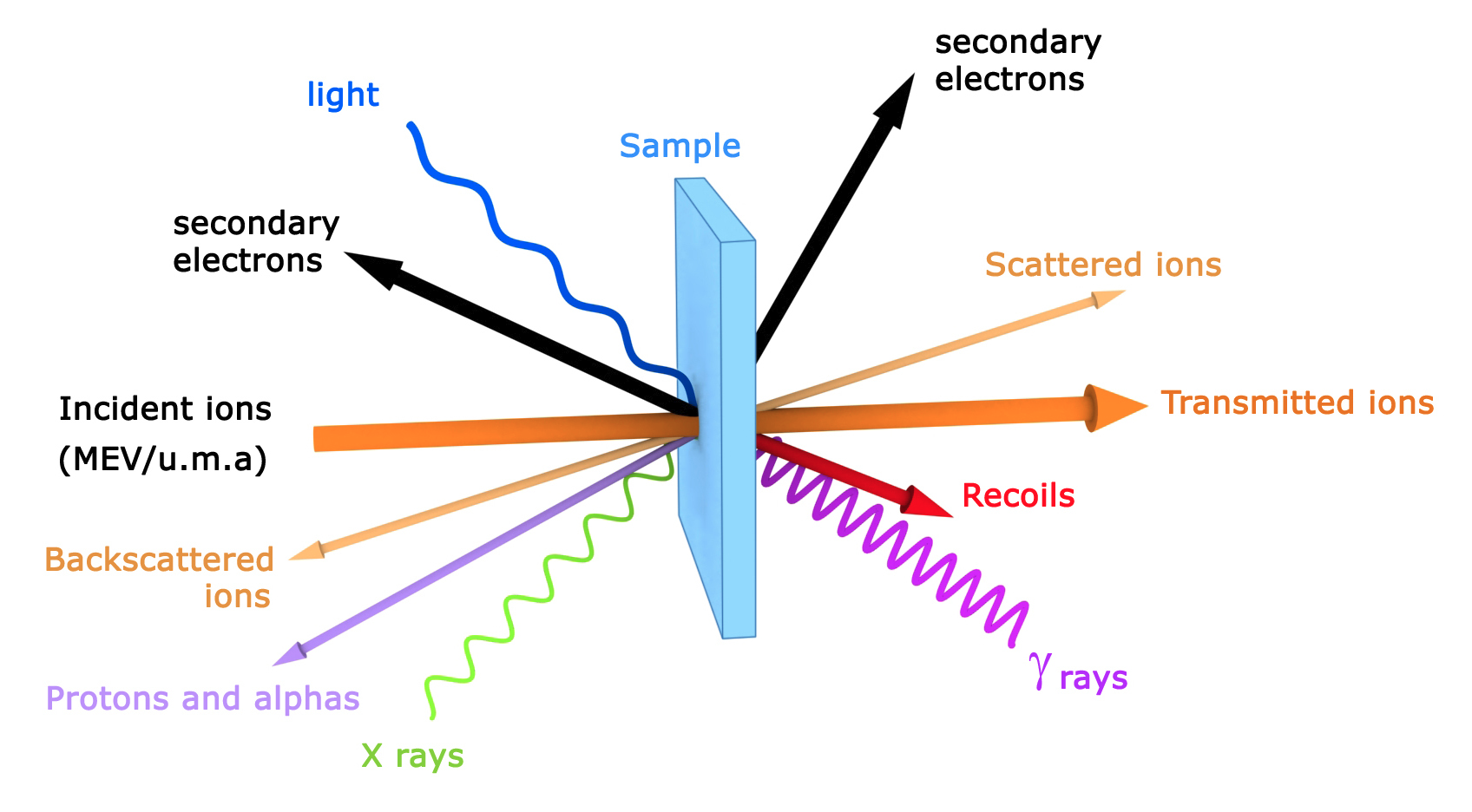}
    \caption{Schematic of MeV-ion beam interaction with matter. Emitted particles may be used for material characterization.}
    \label{fig:schematic}
    \end{figure*}

    The diversity of interaction processes between energetic ions and matter generates a series of signals that can be recorded simultaneously or separately using dedicated instrumentation. These signals provide various types of information about the sample. See the schematic of the signals in Fig. \ref{fig:schematic}. 
    
    The energy spectra of the backscattered ions by the Coulomb nuclear potential form the basis for Rutherford Backscattering Spectrometry (RBS). This technique enables elemental depth profiling with nanometer-scale resolution through energy-loss calculations\cite{jeynes_thin_2016}. When ions approach the nucleus closely enough to interact with the nuclear force potential, resonant enhancements in the interaction cross-section for specific elements can occur, leading to Elastic Back-scattering Spectrometry (EBS). EBS shares the depth-profiling capabilities of RBS, but provides enhanced sensitivity to selected elements\cite{jeynes_book}. 
    
    The collision of the energetic ion may result in an energetic atomic recoil, which can be detected and recorded for a very sensitive characterization of shallow regions close to the surface. This technique is known as Forward Recoil Spectrometry (FRS) or Elastic Recoil Detection Analysis (ERDA)\cite{davies_elastic_1998}.  At even higher ion energies, nuclear reactions may be induced, resulting in the emission of either an energetic photon (Particle-Induced Gamma-ray Emission, PIGE) or a particle such as a proton or alphas (Nuclear Reaction Analysis, NRA)\cite{kokoris_book}. Additionally, when energetic ions interact with atoms in a sample, they can eject inner-shell electrons. As electrons from higher energy levels fill these vacancies, characteristic x-rays are emitted. The Particle-Induced X-ray Emission (PIXE) technique analyzes the energy and intensity of these x-rays to identify and quantify elements with high sensitivity and accuracy\cite{ishii_pixe_2019}. Finally, ion channeling effects can be exploited to obtain structural information about the material\cite{vantomme_50_2016}. 

    These techniques provide IBA with great sensitivity and flexibility for efficient and robust material characterization, with features that match most of the modern constraints for material analysis in diverse fields. In addition, the great power of IBA lies in the simultaneous exploration of more than one of these techniques in a unique model\cite{jeynes_total_2012}, constraining the confidence level and providing robust sample characterization. 
    
    
\subsection{Machine Learning and IBA}
    
\begin{figure*}[htb!]
    \centering
    \includegraphics[width=\textwidth]{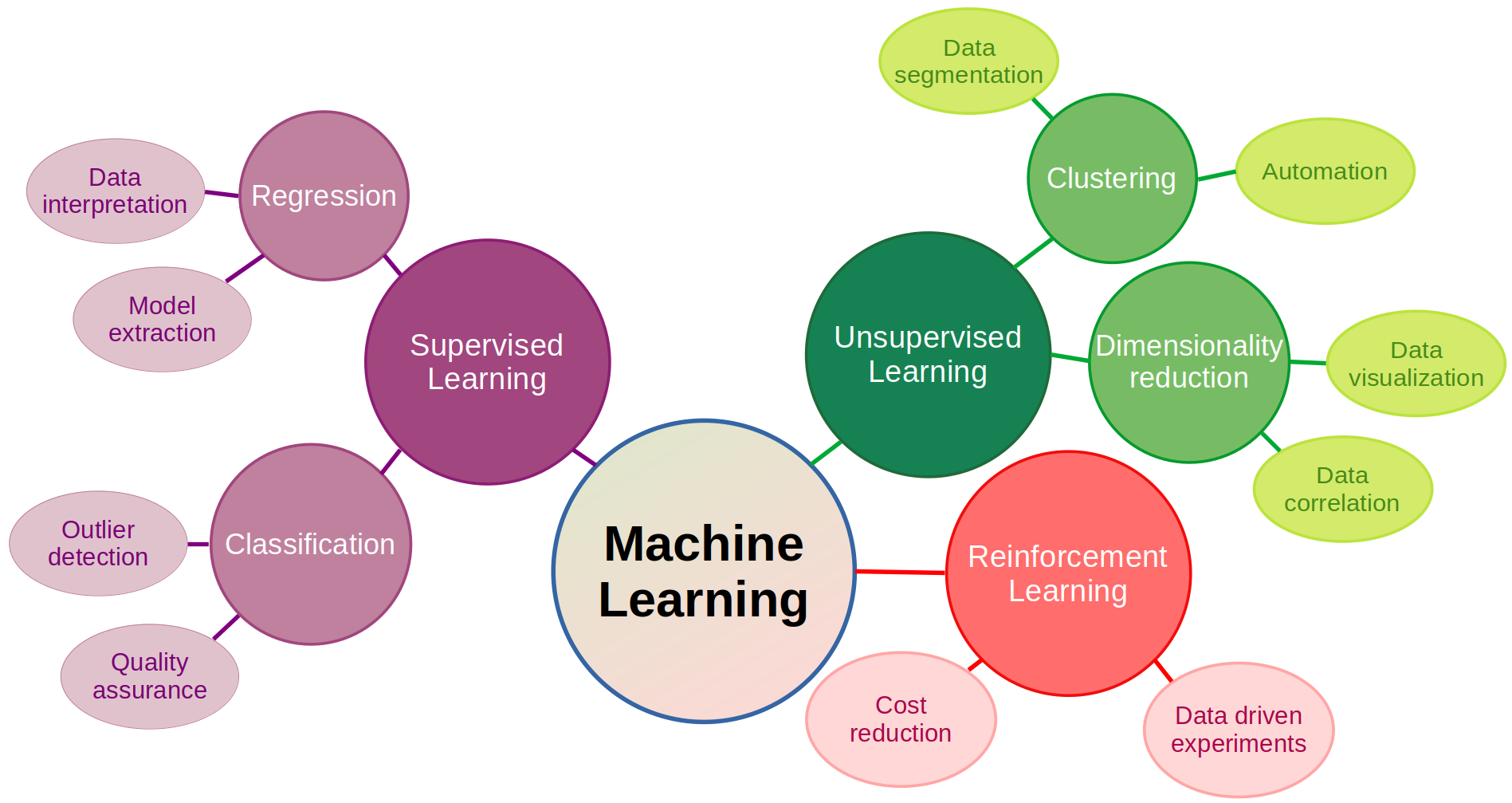}
    \caption{Overview of machine learning algorithms.}
    \label{ML_overview}
    \end{figure*}

    MLA has three main categories of algorithms: supervised learning, unsupervised learning, and reinforcement learning. Each category presents features that enable for some particular tasks to be learned. A schematic of categories and functionalities is summarized in Fig. \ref{ML_overview}. 
    
\begin{figure*}[htb!]
    \centering
    \includegraphics[width=12cm]{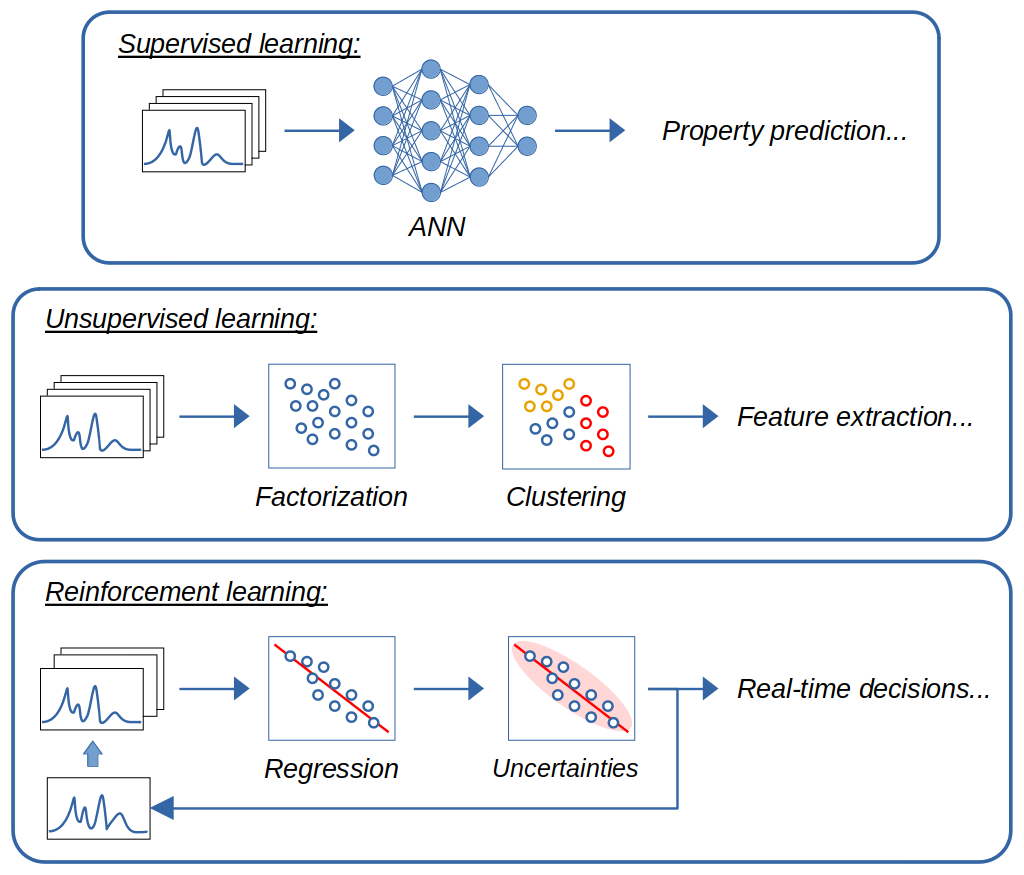}
    \caption{Summary of potential applications of machine learning in IBA facilities.}
    \label{ML_summary}
    \end{figure*}

    Given the characteristics, we may generalize its role in data workflow. For example, supervised learning can predict sample characteristics based on spectral data, while unsupervised learning can extract features from large datasets. Finally, reinforcement learning may find its role in real-time decisions-making in a new type of experiment, driven by incoming data. This is shown schematically in Fig. \ref{ML_summary}. Of course, as with any generalization, this may be subject to exceptions. 
    
    One proposal for adopting MLAs in the data processing workflow consists of implementing layers of automated data processing that act as a virtual assistant in the analysis \cite{silva_ion_2020}. These systems can extract valuable information from the data while reducing the data processing time. 
    
    In the next sections, examples of MLA applied in the IBA field are discussed in terms of their strengths and potentials.

\subsubsection{\label{supervised_ML}Supervised learning}
    
    Supervised learning is a fundamental paradigm in machine learning where the algorithm learns how to make predictions or decisions based on labeled training data. In other words, the algorithm is trained with a data set that includes input features and their corresponding target outputs to learn a mapping that can accurately generalize from the training examples to make predictions about the output of new input data. The training procedure aims to minimize the discrepancy between the model predictions and the true target values, typically by adjusting its internal parameters through special techniques like back-propagation. This approach is widely used in tasks such as image recognition, natural language processing, and predictive modeling, where the algorithm learns to identify patterns and relationships in the data to make informed decisions. The most notable example of supervised learning is Artificial Neural Networks (ANN). They work as universal function approximators, and the architecture of the network offers plasticity and flexibility for a wide range of applications. 
    
    To process data from material characterization with ANN, there are two distinct methods to organize an effective training data set. It may consist of many experimental data on actual reference samples; however, obtaining sufficient data to train a robust ANN using such an approach would require an exceedingly large number of measurements on samples with a wide variety of elements and compositions. Alternatively, one may rely on spectra simulation, which can then be augmented rapidly and systematically to account for possible artifacts (noise, pile-up, background, etc.). This approach is particularly suitable for IBA, where there is a great availability of excellent quality simulation codes for scattering techniques \cite{mayer_computer_2011}. Simulation codes like SIMNRA \cite{mayer_improved_2014, mayer_computer_2017, mayer_multiple_2006} and NDF \cite{barradas_rbs_1998, barradas_advanced_2003, barradas_advanced_2008} are extensively compared against experimental data, and even with some differences in the methodologies, the agreement of the simulated spectra confirms the robustness of the physical models and algorithms \cite{rauhala_status_2006}. Self-consistent analysis \cite{jeynes_total_2012} increases confidence in the quality of simulations even better, since the convergence of multiple techniques to a single sample description depends strongly on the accuracy of the models and fundamental data (such as cross-sections, screening, stopping powers, and straggling) \cite{barradas_unambiguous_1999, jeynes_accuracy_2020}. More recently, simulation software included a virtual assistant to evaluate options and simulation parameters in order to suggest possible improvements to the user in analysis quality \cite{mayer_expert-assisted_2022}. This is a unique scenario among competing analytical techniques \cite{robinson_selecting_2024} that gives confidence in the use of simulation data as the training set. In this sense, producing a large enough dataset of input data and their corresponding output is a matter of computational time. 

    The application of ANNs in the interpretation of RBS spectra had its principle proved in the 2000s for thin film analysis \cite{barradas_artificial_2000, vieira_neural_2000, barradas_artificial_2002}, including evaluation of performance errors \cite{vieira_error_2001} and application on multiple spectra \cite{nene_artificial_2006}. However, a thorough quantitative comparison of performance against conventional methods was performed just recently on a large data set of marker layers for erosion studies of the inner walls of the W7x fusion reactor \cite{guimaraes_processing_2021}. Also recently, the full exploitation of synergies between different IBA techniques using MLA was demonstrated \cite{solis-lerma_automated_2022}.  
    
    The main conclusion of these references is a clear advantage for the adoption of ANNs when the number of experimental data is massive, reducing the distance between IBA measurement and scientific conclusions \cite{demeulemeester_artificial_2010, mayer_material_2020, mayer_carbon_2022}. Another observation is that the training datasets significantly affect the quality of the results: the ANN can be more general if labeled data are widespread in the output space, while it can be more specific if this variation is limited to a certain interval. However, the main pitfall of ANN-based data analysis comes in extrapolation conditions, i.e., when the ANN is consulted on data that lie outside the training region. Even though the ANNs are very good in interpolations, extrapolated predictions tend to fail considerably. A solution for that is the adoption of committee machines, in which the responses of multiple ANNs with different architectures and trained with the same data set are combined. If consensus is reached among the ANNs, the prediction stays within the training region and thus tends to be accurate. Otherwise, none of the trained ANNs can provide good results \cite{ensemble}. 
    
    A feature of IBA that has recently been demonstrated as an important advantage compared to its main competitors is related to accuracy limits and the establishment of a complex traceability chain of the results \cite{jeynes_quality_2006, jeynes_accurate_2012, colaux_high_2014, cureatz_improving_2022}. The adoption of ANNs in the data processing workflow does not necessarily exclude this feature. Although ANNs are complex and pose a large number of parameters, protocols can be created to evaluate the uncertainties inserted by the algorithm \cite{vieira_error_2001, guimaraes_processing_2021, oliveira_what_2022, magchiels_machine_2024, magchiels_enhanced_2024}, and the interpretability of the results can be demonstrated using techniques similar to Average Gradient Outer Products (AGOP) \cite{oliveira_what_2022}. This is an important research topic in the implementation of intelligent data analysis in general but especially for traceable material analysis, and we have been working on this for the specific case of IBA data processing.

    Other ANN architectures have been used to better exploit the potential of nuclear scattering techniques as primary measurement standards. When accurate and traceable measurements are made, the estimation of uncertainties is of high importance. In this sense, some work has searched for alternatives to obtain such an estimate from conventional fully connected multilayer perceptron networks \cite{guimaraes_processing_2021, magchiels_machine_2024}. However, using Mixture Density Networks, the authors demonstrated the principle of modeling the depth profile with the associated uncertainties, embedding the task of data analysis and the evaluation of statistical significance in the same model, all within a Bayesian perspective \cite{muzakka_analysis_2024}.

    The main result of the adoption of ANNs in spectral data processing is the fact that it outperforms the conventional data evaluation procedure in most cases in terms of delivery time or consistency. In the current state of implementation, it still lacks generality and fails in more complex samples. However, it becomes clear that the extension of the applicability of IBA is a direct result of the speed of data processing \cite{mayer_material_2020, mayer_carbon_2022, magchiels_enhanced_2024}.

    It should be mentioned that supervised learning is not exclusively applied to nuclear scattering techniques, nor are ANNs the only tools to map inputs to outputs. In recent work a set of techniques, including PIXE, EBS, PIGE, and RBS, have been jointly processed using Gradient Boosting \cite{xgboost} to analyze air pollution filters \cite{cohen_machine_2024}. The algorithm was trained using data collected for decades and processed using conventional approaches. It can process new data measured under the same experimental conditions, keeping consistency with previous analysis. However, a key point is that standard tools for the interpretation of PIXE data do not feature spectrum simulation \cite{ryan_quantitative_2002, campbell_guelph_2021}. Not many codes provide this feature, and even the initiatives in this direction are not widely available or as intensively tested as the nuclear scattering simulation codes\cite{reis_fixed_2014, ishii_continuous_2006, pixesim}. This fact limits the availability of training data and is the main reason why ANNs find restricted applications in the PIXE data processing workflow.

\subsubsection{\label{unsupervised_ML}Unsupervised learning}
    
    Unsupervised learning is another branch of machine learning in which algorithms are selected to filter, compress, and cluster data without explicit supervision, which means that they work with unlabeled or partially labeled datasets. The primary objective of unsupervised learning is to discover underlying patterns, structures, or relationships within the data without any predefined target outcomes. It often involves statistically sound techniques to account for correlations and density estimation, with applications ranging from grouping similar data points to reducing the complexity of data for visualization or further analysis. Unsupervised learning is particularly valuable for tasks such as data segmentation, anomaly detection, and data exploration to reveal hidden correlations that may not be apparent through manual labeling or guidance. All of these tasks are particularly interesting in material analysis when applied to experimental data in a large number of samples or images.

    It is also common that, in a data process workflow, unsupervised learning is used to reduce the dimensionality of the data before some supervised learning algorithm. It is a response to two different issues: i) the unsupervised learning will filter out the random noise and group together correlated data; ii) by the data reduction the supervised learning models can be simplified (less hidden layers in an ANN, for example) avoiding the "curse of dimensionality". Here, an interesting feature of MLA is revealed: the algorithms can be applied systematically as in a layered stack, with the input of one algorithm being the output of the previous one. Therefore, a complex chain of algorithms can be organized into an efficient data processing workflow.

    Statistical inference and multivariate analysis are now included as topics in the unsupervised learning branch of MLA due to their ability to reveal hidden patterns in the data. Multivariate analysis has been applied in IBA for a long time, mainly in PIXE studies on a large number of samples \cite{martins_chronological_1999,bernardes_ascertaining_2014}. Currently, it has been used with a different focus assisting in the development of imaging capabilities of IBA. Whether on micro-\cite{giuntini_review_2011} or macroscopic scales \cite{silva_elemental_2018}, IBA has been proven to be an efficient tool for producing hyperspectral images of samples. Experimental setups with multiple detectors provide a multitechnique approach in single irradiation, but PIXE seems to be the preferred technique in the mapping modality of IBA techniques. Initially, the adequacy of conventional data processing for images included increasing computational processing capabilities \cite{PICHON20102028, PICHON201548}. More recently, the data processing approach has included an additional layer in the workflow that exploits unsupervised algorithms to enable image segmentation \cite{doyle_pixe-quantified_2006, silva_multivariate_2018, mihalic_multivariate_2021, tazzioli_advanced_2024}. A similar approach has been adopted for x-ray fluorescence using synchrotron radiation \cite{de_la_rosa_clustering_2024} and atomic force microscopy \cite{rahman_laskar_scanning_2023} data, showing a common interest with other communities of material characterization.

    Here again, MLA extends the applicability of IBA since the image segmentation enables similar pixels to be summed together by similarities (clustering), which brings two benefits: 1- a dramatic enhancement of the lower quantification limit in mapping modality (sensitivity); 2- provides better statistical representation of selected areas. By applying dimensionality reduction techniques such as nonnegative matrix factorization (NMF), correlations in signals of different elements can be exploited to extract the fluorescence spectra of compounds from raw IBA data of decorative tiles \cite{silva_multivariate_2018} or paintings \cite{mihalic_multivariate_2021, tazzioli_advanced_2024}. The compound information is hardly reachable by IBA techniques, and exploiting spatial correlation seems to be a possible way to obtain this information. Lastly, clustering pixels with similar compounds results in an enhanced sensitivity compared to single-pixel analysis.

    Utilizing unsupervised algorithms as a preprocessing layer before the use of conventional tools for IBA spectral processing seems to be a promising way to save computational time and extend image interpretability. In this approach, full physics simulations are performed only to interpret some spectra obtained as MLA results, significantly reducing the need for large computational resources and, at the same time, processing data with much more statistical significance \cite{tazzioli_advanced_2024}. All of this results in a gain in qualitative information and a reduced demand for computational power to obtain quantitative information. 

\subsubsection{\label{reinforcement_ML}Reinforcement learning}
    
    Reinforcement learning is a machine learning paradigm in which an agent interacts with an environment and learns to make sequential decisions to maximize a cumulative reward. Unlike supervised learning, it does not rely on explicit data labels but instead involves trial and error. The agent takes actions in the environment and receives feedback in the form of rewards or punishments, which it uses to refine its decision-making policy. The ultimate goal is to discover a policy that, over time, optimally balances exploration (trying new actions to learn) and exploitation (choosing known good actions) to achieve long-term goals. Reinforcement learning finds applications in diverse fields, such as robotics, autonomous systems, and game play, where it enables agents to learn and adapt to complex, dynamic environments.

    Bayesian optimization is a general methodology that is often utilized when each data point is expensive to acquire in terms of computational time or experimental efforts. It has been proven efficient in IBA as a tool for experimental design \cite{von_toussaint_optimizing_2010, schmid_statistically_2012, silva_bias_2022} to extend the applicability or optimize the information. Although in these references the targeted optimization is done offline, it can be implemented in an online protocol to guide the sequence of data acquisition in a multistep experiment, varying experimental conditions such as ion energy, scattering or incidence angle, or any other experimental setup parameter that may influence the analysis outcome (depth profiling resolution, sensitivity, etc.). Optimizations are usually obtained through the calculation of activation functions that aim to maximize the expected improvement with a new measurement \cite{silva_bias_2022} or minimize the general variance of the analysis results.

    The challenge in the implementation of reinforcement learning allied to experimental design is the computational costs involved in the calculation of some performance metrics in the framework of information theory, such as the Kullback-Leibler divergence (information gain). This challenge comes from the fact that a large number of physical simulations are required. This can be accelerated by parallel computing, but further developments are still required.  

\section{Perspectives and future pathways}

    The field of MLA is rapidly evolving and presents changes and breakthroughs quite often. Generative models are currently in fashion and show surprisingly good results in image, video, and audio generation. In IBA, we can take advantage of this technology to solve complex and computationally demanding problems. An example is the analysis of laterally inhomogeneous samples, which demands specific simulation codes that are highly complex and present time-consuming calculations \cite{mayer_computer_2016, mayer_computer_2017, mayer_intercomparison_2016, claessens_quantification_2022}. Implementing conventional optimization algorithms is feasible; however, the computational time for convergence is prohibitive. A generative model that can learn patterns from simulations can replace simulation software and speed up calculations. A similar approach can benefit the adoption of reinforcement learning to optimize experiments.

    The traceability of the uncertainty budget has open points if one considers the complexity of simulations. Significant progress has been made in the traceability of RBS \cite{colaux_certified_2015} and in the development of a set of computational tools to estimate uncertainties \cite{barradas_bayesian_1999, pascual-izarra_simultaneous_2006}. However, the uncertainty of the fundamental databases cannot be easily propagated, and difficulties are also present when considering geometrical uncertainties in non-Rutherford conditions. Thus, taking into account the work done with other techniques that present similar difficulties, machine learning algorithms can be used to tackle this problem. For example, an approach based on Bayesian deep learning has been used to quantify uncertainties in the inverse problem of x-ray scattering data interpretation~\cite{yang_uncertainty_2024}, outperforming the standard Markov-Chain Monte Carlo (MCMC) approach by three orders of magnitude in time.

    Another example is the Large Language Models (LLMs), with many commercially available options. This technology can massively process countless amounts of research articles and present correlations, trends, and solutions to many open questions. Similar initiatives are already emerging in the field of material science that have excellent results \cite{boiko_autonomous_2023}. LLMs can, for instance, replace traditional graphical user interfaces with natural language prompts, making complex tools more intuitive and accessible, particularly for nonexperts. In addition, they can automate report generation by transforming raw data into polished, domain-specific documents tailored to technical or general audiences. LLMs are also highly effective at incorporating non-quantitative information into analysis workflows. They can interpret and integrate textual data - such as research articles, books, experimental logs, and qualitative observations - alongside numerical data. This capability allows for the inclusion of contextual details, such as descriptions of sample preparation processes, into decision making and result interpretation. By combining qualitative and quantitative insights, LLMs can deliver context-rich analyses, identify anomalies, and suggest hypotheses for unexpected findings. Although challenges such as data privacy, accuracy, and integration with legacy systems persist, LLMs have the potential to revolutionize material analysis by bridging gaps between disciplines and enabling more efficient and insightful research.

    Multimodal data integration is a key focus in the development of MLAs. IBA techniques provide quantitative data on elemental composition, while complementary methods such as electron microscopy, X-ray Diffraction (XRD), and spectroscopy yield structural, morphological, or molecular-level insights. MLAs can integrate these diverse datasets, uncovering correlations and complementarities that are difficult to identify manually.

    One of the primary challenges in multimodal data integration is image alignment, as datasets often differ in physical properties such as scale, color depth, and resolution. Addressing this requires complex transformation models, where MLAs can play a crucial role by detecting keypoints, identifying landmarks, or analyzing intensity similarities. Algorithms may either directly perform the necessary transformations or learn feature representations to facilitate alignment.

    Dimensionality reduction techniques, such as Principal Component Analysis (PCA) and t-distributed Stochastic Neighbor Embedding (t-SNE), are invaluable in this process. They condense high-dimensional data from multiple methods, preserving essential features while enabling intuitive visualization and analysis. In addition, MLAs can evaluate measurement uncertainties and resolve discrepancies between data sets obtained from different techniques, which can be explained by the range of uncertainty coverage, thereby enhancing the reliability and robustness of conclusions.
    
    Looking into a more fundamental perspective, there is also the Physical-Instructed Neural Networks (PINNs). In this case, one may use physical models and boundary conditions to write a customized loss function or constraints to ANNs optimization, making the necessity of a training dataset optional. In this case, the ANN can solve complex physical problems based on basic physics principles. This can be particularly useful, for example, in the development of stopping power models of ions in matter. Some initiatives to improve stopping power data tables using machine learning algorithms are based solely on training models with experimental data \cite{parfitt_machine_2020, bivort_haiek_espnn_2022, akbari_predicting_2023}. However, these approaches face complex challenges in data cleansing and a lack of physical constraints. The first challenge is related to problems associated with an experimental measurement data set, including systematic errors, poor uncertainty estimation, and difficulties associated with the selection of which data to trust and what data should be discarded. The second challenge is related to restricting model predictions into physics theories constrained boundaries, thus avoiding ambiguities and discontinuities, and, in the last instance, dealing with overfitting. Because a complete theory that is capable of accurately dealing with different regimes of energy-loss processes is not available at the moment, imposing these constraints in the models seems difficult. However, some new theoretical approaches seem promising in that sense \cite{matias_efficient_2024}.
    
\section{Discussion and conclusions}

    In the previous sections, we have seen how the three branches of MLA can be incorporated into scientific work in the IBA field, potentially facilitating advancements.

    A favorable environment for the adoption of MLA is formed by the excellent availability of simulation software (forward models) together with a good understanding of the physical processes involved. This fact facilitates the production of training datasets and enables the use of large ML models. However, there is a need for an efficient data workflow using multiple detectors and the mapping modality of IBA that produces a large amount of experimental data to process and interpret. 
    
    Even in this favorable scenario with important development demands, the adoption of MLA is still limited in IBA. One possible reason may be the lack of standardized tools with a convenient user interface to facilitate access to all users. We have seen a similar situation in the past with the spread of IBA analysis with the advent of more friendly simulation software. Currently, what is demonstrated in the literature about the potential use of MLA in IBA is still done at a low level of coding, restricting access to skilled programmers. We can also include the lack of standardization of data format files as a reason for not spreading MLA since it could enable data and trained model exchange. As a result, the tools developed are still restricted to the research group that developed it.

    It is now clear that conventional approaches and protocols cause delays in IBA throughput and analysis delivery, imposing an important drawback to IBA in front of its competitors. At the same time, the incorporation of MLA to speed up the analysis workflow must be accompanied by the development of validation and verification tools, uncertainty analysis, and traceability, in order to keep IBA competitive without losing its reliability.
    
    There is still room for improvement, either in innovative applications, enhancing workflows, or obtaining new insights. Reinforcement learning can be an important component in next-generation experimental setups, resulting in improved reliability and extending applicability. 

    Physics-informed models are still under development but already promise an efficient method to obtain insights from complex physical phenomena, providing theory-consistent conditions to MLA.

\begin{acknowledgments}
    The author thanks the financial support of CNPq-INCT-FNA (project number 464898/2014-5), CNPq (project number 406982/2021-0) and FAPESP (project number 2022/03043-1).
\end{acknowledgments}

\section*{Author Declarations}
The authors have no conflicts of interest to disclose.

\section*{Data Availability Statement}
Data sharing is not applicable to this article, as no new data was created or analyzed in this study.

\bibliography{aipsamp}

\end{document}